\documentclass[usenatbib, fleqn]{mnras}
\usepackage{amsfonts}
\usepackage{amsmath}
\usepackage{amssymb}

\usepackage[T1]{fontenc}

\usepackage[dvips]{graphicx}
\usepackage{myaasmacros}
\usepackage{color}

\def\bea{\begin{eqnarray}}
\def\ena{\end{eqnarray}}

\def\acta{Acta Astron.}

\def\pdot {\dot P}

\def\msun{{\rm M}_{\odot}}
\def\rsun{R_{\odot}}

\def\mdot {\dot M_{\rm W}}

\def\ltsima{$\; \buildrel < \over \sim \;$}
\def\lsim{\lower.5ex\hbox{\ltsima}}
\def\gtsima{$\; \buildrel > \over \sim \;$}
\def\gsim{\lower.5ex\hbox{\gtsima}}
\def\hd {HD\,49798}

\def\hr {HD\,49798/RX\,J0648.0--4418}

\newcommand{\ms}{\mbox {$M_{\odot}$}}

\def\apgt{{\raise-.5ex\hbox{$\buildrel>\over\sim\,$}}}
\def\aplt{{\raise-.5ex\hbox{$\buildrel<\over\sim\,$}}}

\title[A young contracting  white dwarf in the peculiar  binary \hr\ ?]
{A young contracting  white dwarf in the peculiar  binary \hr\ ?}  

\author[S.~B.~Popov,  S. Mereghetti, S.I.~Blinnikov, A.G. Kuranov,  L.R. Yungelson]{ S.~B.~Popov$^{1}$\thanks{Corresponding author. E-mail: polar@sai.msu.ru}, S.~Mereghetti,$^{2}$
\thanks{E-mail:sandro@iasf-milano.inaf.it} 
S.~I.~Blinnikov,$^{1,3,4}$
\thanks{E-mail:Sergei.Blinnikov@itep.ru} 
A.~G.~Kuranov, $^{1}$
\thanks{E-mail: alexandre.kuranov@gmail.com}  
L.R. Yungelson$^{5}$
\thanks{E-mail: lev.yungelson@gmail.com}
\\
$^{1}$ Sternberg Astronomical Institute, Lomonosov Moscow State University, Universitetsky prospekt 13, 119234, Moscow, Russia\\
$^{2}$ INAF, IASF-Milano, Via E. Bassini 15, I-20133 Milano, Italy\\
$^{3}$ ITEP, B.Cheremushkinskaya 25, 117218, Moscow, Russia\\
$^{4}$ Kavli IPMU, Kashiwa, Japan\\
$^{5}$ Institute of Astronomy, Russian Academy of Science, 48 Pyatnitskaya Str., 119017, Moscow, Russia\\
}

\date{Accepted, Received}

\pubyear{2017}

\begin{document}

\maketitle


\begin{abstract}
\hr\ is a peculiar X-ray binary with a hot subdwarf 
(sdO) mass donor. The nature of the accreting compact object is not known, but its  spin period $P=13.2$~s and $\dot P =-2.15 \times 10^{-15}$s~s$^{-1}$, prove that it can be only either a  white dwarf or a neutron star. The spin-up has been very stable for more than 20 years.  
We demonstrate that the continuous stable spin-up of the compact companion of HD 49798 can be best explained by contraction of a young white dwarf with an age $\sim 2$~Myrs.  
This allows us to interpret all the basic parameters of the system in the framework of an accreting white dwarf. We present examples of binary evolution which result in such systems. If correct, this is the first direct evidence for a white dwarf contraction on early evolutionary stages.
\end{abstract}


\section{Introduction}
\label{sec:intro}

\hr\ is a peculiar binary consisting of an X-ray pulsar, with spin period $P=13.2$~s, and a hot subdwarf of O spectral type in a circular orbit with period $P_\mathrm{orb}=1.55$~days
\citep{1970MNRAS.150..215T,1994Obs...114...41S,1997ApJ...474L..53I, 2011ApJ...737...51M}. It is the only confirmed X-ray binary with a hot subdwarf mass donor.
In fact, its X-ray emission is most likely powered by accretion of matter from the weak wind of the sdO star \hd\ (mass loss rate $\mdot=3\times10^{-9}$ $\msun$ yr$^{-1}$,  \citealt{2010Ap&SS.329..151H}), although it is still unclear whether the accreting object is a white dwarf (WD) or a neutron star (NS). 
Its mass is well constrained by a dynamical measurement yielding $1.28\pm0.05\, M_\odot$ \citep{2009Sci...325.1222M}, which fits well both possibilities. 
The evolution of this system was recently studied by \cite{2017arXiv170806798B}.
  
The relatively low value of X-ray luminosity $L_\mathrm{X}\sim 2\times 10^{32}\, (d/650\, \mathrm{pc})^2$~erg~s$^{-1}$,  as well as the X-ray spectrum (a very soft blackbody of temperature $kT\sim$30 eV and large emitting radius $R\sim$40 km plus a hard power-law  tail)  favoured a  WD interpretation \citep{2009Sci...325.1222M,2011ApJ...737...51M}. 

However, recently \cite{2016MNRAS.458.3523M} were able to measure for the first time the secular evolution of the spin period by phase-connecting all the available X-ray observations spanning more than 20 years. They discovered that the compact companion of \hd\ is spinning-up  at a rate  $\pdot = 2.15\times 10^{-15}$~s~s$^{-1}$ (the period derivative is negative, but here and everywhere below we refer to its absolute value). 

In the framework of accretion-driven spin-up it is difficult to explain such a  high $\pdot$ value for a WD.
In fact, as shown in \citet{2016MNRAS.458.3523M}, this would require that \hd\ be farther than $\sim4$ kpc, a distance inconsistent with  that derived from optical/UV studies (650$\pm$100 pc, \citealt{1978A&A....70..653K}). A NS, thanks to its $\sim10^5$ times smaller moment of inertia, seems less problematic. However, also in the NS case,  some puzzles remain, such as, e.g., the large emitting area of the blackbody component,  the extremely steady luminosity and spin-up rate over more than 20 yrs which are quite unusual in wind-accreting neutron stars, and the requirement of a NS magnetic field lower than $3\times10^{10}$~G to avoid the propeller effect  \citep{2016MNRAS.458.3523M}.
 
In this paper we propose a completely different  explanation for the spin-up of the compact companion of \hd , not related to accretion. We propose that the object is a young WD, still contracting and thus with a decreasing moment of inertia. In the next section we describe the model we used to calculate the WD evolution. In Sec. 3 our results are presented and the age estimate for the WD is provided. 
In Sec. 4 we discuss our hypothesis and, finally, we conclude in Sec. 5.
Through the paper we use the notation $N_\mathrm{X}=N/10^X$.

\section{Model of white dwarf evolution}

Theories of the WD evolution predict the dependence of luminosity $L$ and effective temperature 
$T_{\rm eff}$ on their age $t$.
White dwarfs belong to the old population
of the Galactic disk, so their birthrate for the last billion years remains
constant and their number within few tens of parsec from the Sun does not depend on
our position relative to the spiral arms. That is why the number of   WDs 
per unit volume in a luminosity (or $T_{\rm eff}$) interval  
is proportional to the time they spend in this interval. This allows one to check the validity
of the theory of WD evolution.

To calculate an evolutionary sequence of a WD we apply the code developed by  \cite{1993ARep...37..187B, 1994MNRAS.266..289B}. 
The modeling of the WD evolution is done with account of the
data on the electron heat conductivity \citep{1980SvA....24..126U,1980SvA....24..303Y,1983ApJ...273..774I,1984MNRAS.209..511N}, 
 the rate of neutrino losses \citep{1963PhRv..129.1383A, 1967ApJ...150..979B,
 1972PhRvD...6..941D,1985ApJ...296..197M, 1989ApJ...339..354I},
the equation of state \citep{1996ApJS..106..171B,1989ASPRv...7..311Y}
and Coulomb
screening \citep{1989ASPRv...7..311Y} 
in thermonuclear reactions for the hot dense plasma. 

To start the evolution for a given WD mass we construct 
an artificial hydrostatic model which first allows us to
get a hot WD with $T_{\rm eff} > 10^5$~K, and this WD  later cools down. 
We have constructed a hydrostatic configuration by the method of \cite{1986NInfo..61Q..29N} 
for  calculating the initial model with a much larger radius than that of the WD. We begin our runs using  an implicit hydrodynamic solver, and the stellar model is quickly heated up to 
$T_{\rm eff} > 10^5$~K by the influence of gravitation. 
One can start to compare the results with the observations from the moment at which the WD cools down back to
$T_{\rm eff} \approx 10^5$~K and the initial conditions become inessential.

Our cooling curves reproduce quite well the observed luminosity function of WDs.
Moreover, since the cooling curves of hot WDs are sensitive to the WD mass,  $M_{\rm WD}$, due to sensitivity of plasma neutrino emission to density, \cite{1994MNRAS.266..289B} were able to derive the best fitting mean WD mass, 
which was found to be higher than the value of $\sim$0.55~$M_\odot$   usually adopted at that time \citep{2008AJ....135.1225H}.  
However, the most recent measurements of the average WD mass are in good agreement with the original predictions of \citet{1994MNRAS.266..289B}.
A mean mass, $\langle M_{\rm WD} \rangle = 0.642 M_\odot$, is found in the full 25~pc WD sample by \cite{2016MNRAS.462.2295H}.
This can be compared with 
a mean mass of $\langle M_{\rm WD} \rangle = 0.650M_\odot$  found by \cite{2012ApJS..199...29G} in a 20 pc
sample, and  $\langle M_{\rm WD} \rangle = 0.699 M_\odot$ published by \cite{2015ApJS..219...19L} for their 40 pc sample.
Thus, the mean WD mass published by \cite{1994MNRAS.266..289B} two decades before
these modern estimates can be considered as a blind test of the correctness of their code used in the current paper. 
 In the Appendix, we present a more detailed comparison of our WD evolution code with two modern codes, based on the  results obtained for different calculations in $L-t$ and $T_{\rm eff}-t$ plots.

\begin{figure}
\centering
\includegraphics[width=\linewidth]{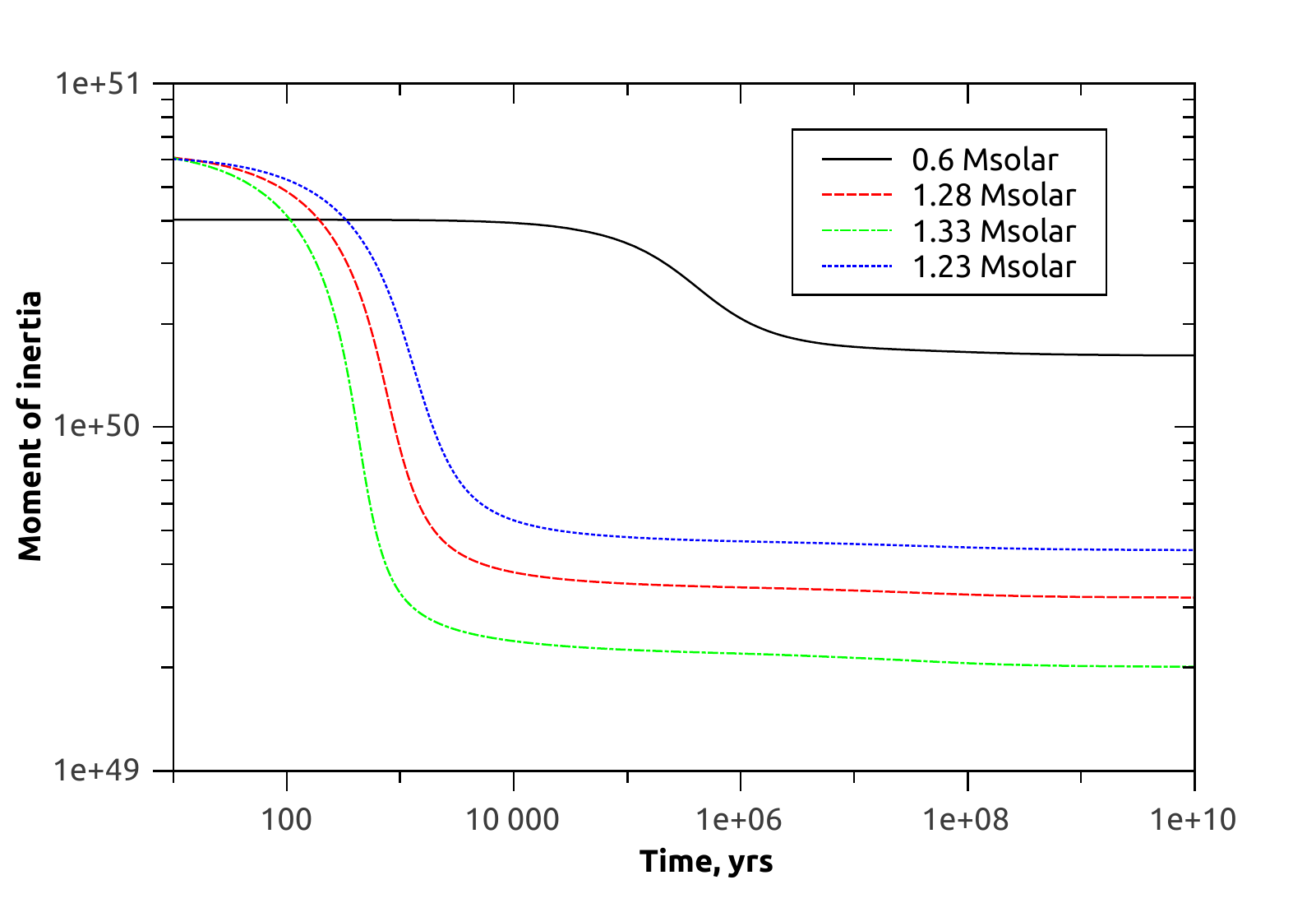}
\caption{Evolution of moment of inertia for four WD masses: $0.6\, M_\odot$ (black solid line), and three values representing the mass range for HD 49798 $1.28\pm 0.05\, M_\odot$ (dashed, dotted, and dash-dotted color lines).  }
\label{fig_momi}
\end{figure}

The WD moment of inertia $I$   at each time step is calculated according to:
\begin{equation}
I=\frac{8\pi}{3} \int_0^R \rho r^4 dr,
\end{equation}
where $R$ is the WD radius. The evolution of $I$ for four values of WD masses is presented in Fig.~\ref{fig_momi}.

\section{Results}

\subsection{Period derivative}
\label{sec_pdot}

We use angular momentum conservation to derive $\pdot$ from the calculated evolution of $I$:


\begin{equation}
I_1/I_2 = P_1/P_2.
\end{equation}
Here all values correspond to two times separated by an interval $\Delta t$.
As the WD is contracting, $I_1>I_2, P_1>P_2.$ So, $P_1=P_2+\Delta P$.
And finally,

\begin{equation}
\dot P = \Delta P/\Delta t= \frac{P}{\Delta t}\left(\frac{I_1}{I_2}-1\right).
\label{pdot_eq}
\end{equation}
Here we use the observed value $P=13.2$~s and moments of inertia are taken from the  evolutionary sequences described above.

\begin{figure}
\centering
\includegraphics[width=\linewidth]{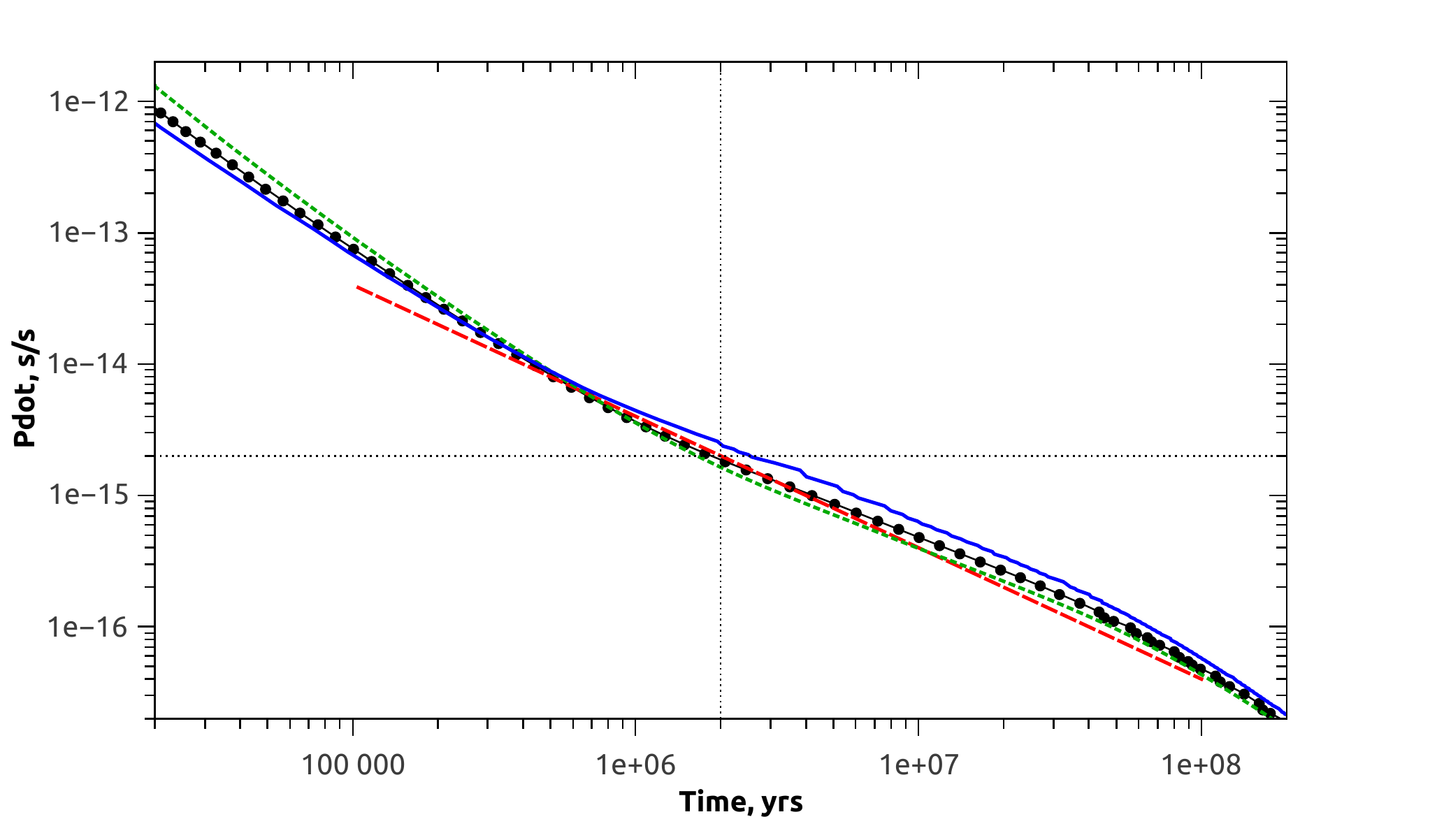}
\caption{Evolution of $\pdot$ according to calculations using eq. (\ref{pdot_eq}). Symbols correspond to the mass $1.28\, M_\odot$. The solid line which initially goes below symbols, and then above, corresponds to $1.33\, M_\odot$. The dotted line which initially goes above, and then below symbols is calculated for $1.23\, M_\odot$.
The dashed diagonal line corresponds to equation $\pdot \propto t^{-1}$. The horizontal dotted line corresponds to the measured $\pdot$ of HD 49798 (Mereghetti et al. 2016) and the vertical dotted line corresponds to the age $2\times 10^6$~yrs. }
\label{fig_pdot}
\end{figure}

We present our results for the evolution of the period derivative in Fig.~\ref{fig_pdot}, where the curves refer to three values corresponding to the uncertainties in the mass of the WD in HD 49798: $M=1.28\pm 0.05\, M_\odot$ \citep{2009Sci...325.1222M}. 

In Fig.~\ref{fig_pdot} we also added the line $\pdot  = 2\times 10^{-15} \, (t/2\times 10^6 \mathrm{yrs})^{-1} $~s~s~$^{-1}$, which fits well the behavior of $\pdot $ in the present epoch according to our model.
This simple analytical fit allows us to estimate the second derivative of the spin period, which in this case is expected to be $ \ddot P \approx 3 \times 10^{-29}\, ({t}/{2\times 10^6\, \mathrm{yrs}})^{-2}$~s~s$^{-2}$. 
Unfortunately, the measurement of such a small value is behind the possibility of the current and near future observations.

The above results show that the observed $\pdot$ can be totally explained by the WD contraction. In principle, some additional spin-up could be provided by accretion. 
To quantitatively evaluate such a possible contribution,  we consider the most favorable case for transfer of angular momentum, i.e. accretion through a disk with inner radius truncated close to the corotation radius $R_\mathrm{co} = (G M P^2/4 \pi^2)^{1/3} = 9\times10^8$ cm.  In this case we have: 

\begin{equation}
2 \pi I \dot \nu = \dot M (G M R_\mathrm{co})^{1/2}.
\label{nudot}
\end{equation}

\noindent
and using $\dot M = L(GM/R)^{-1} = 1.8\times10^{14} L_{32}$~g~s$^{-1}$  we obtain $\dot \nu = 1.1\times10^{-19} L_{32} I_{50}^{-1} (R/3000\,\mathrm{km})^{-1} $~Hz~s$^{-1}$.   This corresponds to a spin-up rate $\pdot = 2\times10^{-17} L_{32}$~s~s$^{-1}$,  which is two orders of magnitude smaller than the measured value. 
%
%
In the case of wind accretion, the expected  spin-up rate due to infalling matter would be even lower \citep{2016MNRAS.458.3523M}.
We  finally note that the measured $\pdot$ has been very stable during the time of X-ray observations (more than 20 years), a situation that has never been observed in  X-ray binaries where the period evolution is driven by the interaction of the rotating object with the mass accretion flow.  So, we conclude that the measured spin-up is driven  by the decreasing moment of inertia of the WD.

As we know the mass of the compact object and its $\pdot$, we can estimate its age and other parameters from the evolutionary sequence. The age is about 2 Myrs for M=$1.28\, M_\odot$ and about 3~Myrs for the  upper value of the mass uncertainty (see Fig.~\ref{fig_pdot}).  
Taking into account possible uncertainties of the WD model, we can conservatively estimate the age range for the WD as $1<\mathrm{Age}<5$~Myrs. 

A WD with $M=1.28\, M_\odot$ might have effective surface temperature $\sim 75,000$~K and luminosity $\sim 0.65\, L_\odot$. The radius is $\sim 3340$~km. 
These values do not contradict the observed properties of the system. 
In fact the optical/UV emission of \hr\ is dominated by the  flux coming from the much larger  sdO star, which has an effective temperature $\sim 47,000$~K  and a luminosity  $\sim 10^4\, L_\odot$.

\subsection{Binary evolution}

The evolution of the HD 49798 binary system has been very recently studied  by \cite{2017arXiv170806798B}. These authors considered both possibilities, a WD or a NS, and concentrated mainly  on the future evolution. In this subsection we discuss the origin of the present day appearance of this binary.

Though both components of \hd\ have rather extreme masses,
the origin of this star may be well understood within the paradigm of formation of 
hot subdwarfs
in close binaries due to stable (via Roche-lobe overflow) or unstable (via common envelope)
mass loss \citep{1976ApJ...204..488M,ty90}. Formation channels for helium subdwarfs accompanied by WDs and detailed models of their population were computed, e.g., by   
\citet{hpmmi02,hpmm03,yt05}. 
These models reproduce the bulk population with ``canonical'' mass of subdwarfs close to 
0.5\,\ms\ and predict, as well, the existence of a ``tail'' of massive ($\apgt 1$ \ms) objects.

A numerical example of an evolutionary scenario for the formation of a binary with parameters rather
similar to \hd\ is presented in  Table~\ref{tab:scen}.
We applied for the modeling the binary population synthesis code BSE 
\citep[][September 2004 version]{htp02}. The crucial parameter of close binaries  evolution,
the efficiency   $\alpha_{ce}$ with which the common envelope is ejected,  was set to 2, while the 
binding energy 
of the donor envelope 
parameter $ \lambda$ was varied depending on the evolutionary stage of the star
as prescribed in the BSE code.
 The choice of $\alpha_{ce}$ is justified by the circumstance that 
its high
value allowed us to reasonably reproduce the Galactic SNe~Ia rate by the  ``double-degenerate''
scenario, as well as  the observed delay time distribution for SNe~Ia \citep{2017MNRAS.464.1607Y}.

\begin{table}     
\caption{Scenario of formation of a binary
similar to \hd. 
Evolutionary stages of stars are abbreviated as follows: MS -- main-sequence, ZAMS -- zero-age MS, RG -- red giant, 
CHB -- central He burning, EAGB and TPAGB -- early and thermally-pulsing AGB stages, respectively, 
CE -- common envelope, He$\star$ - naked helium star (He subdwarf), HeG -- helium giant.   }
\centering
   \begin{tabular}{rccr|l}
\hline
    Time  &   $M_1$   &   $M_2$  &  Period  & Stage \\     
    (Myr)  &   (\ms)  &   (\ms) &  (days)  &   \\     
\hline
     0.0     & 7.0          &  6.75        &     4550.3 &  ZAMS  \\  
    48.8     & 7.06          &  6.75        &      4550.3  &   RG+MS \\
    49.0     & 7.05          &  6.75       &       4551.6 &   CHB+MS \\
	53.0 	 &    6.89  	&  6.75 &    4621.7 &      CHB+RG \\
	53.1     &   6.89   	&   6.75  &     4623.4    &    CHB+CHB    \\ 
	55.0 	&     6.84&     6.69 &     4691.9  &    EAGB+CHB  \\ 
	55.3  &   6.8  &   6.69  &    4657.4    &    TPAGB+CHB  \\ 
	55.7  &   5.96	 &    6.84 &     4101.8    &    CE   \\
\hline
	55.7 &    1.28  &   1.47 &        1.48     &    ONe WD+He$\star$ \\  
	64.1  &   1.28 &    1.43 &      1.52         &     ONe WD+HeG \\
\hline
	64.8  &   1.28 &    1.42&       1.53   &    CE \\ 
	64.8 &   1.28 &    0.83&         0.15      &     ONe+CO WDs \\
	467.5   &  1.28 &    0.83&         0.0004        &   Merger \\
\hline
\end{tabular}       
\label{tab:scen} 
\end{table}

Initially, the binary is rather wide and the primary overflows its  Roche-lobe in the TPAGB stage (see Table~\ref{tab:scen} for the explanation of abbreviations related to different stages of stellar evolution).
Since in this stage the star has a deep convective envelope, the mass loss is unstable and a common envelope engulfing both components forms. The ejection of the common envelope results in the formation of an oxygen-neon (ONe) WD accompanied  by a hydrogen-envelope-devoid star that burns helium in the core and may be observed as a hot helium subdwarf. 
Thus, the  birth of the WD and the He-star 
is simultaneous. Helium stars more massive than 0.8\,\ms\ expand 
after
the core He-burning stage and then turn into ``helium giants'' \citep{pacz71}. According to its position 
in the Hertzsprung-Russell diagram,  the \hd\ subdwarf  is, likely, just in the ``transition'' stage.
Lifetimes of massive He-stars are extremely short and commensurate with the expected age of \hd.
 The model predicts that, after Roche-lobe overflow 
by the expanding He-giant, a second phase of common envelope will occur. After ejection of the latter, 
a pair of ONe and CO WDs will be born, which will merge due to the angular momentum loss via gravitational waves radiation in about 1~Gyr. 
The outcome of such mergers still awaits further studies.  

The scenario shown in  Table~\ref{tab:scen}  does not  reproduce exactly the parameters of \hd, but our only goal was  just to show the viability of formation of \hd-like systems. Of course, a better agreement may be obtained by fine tuning of, e.g., common envelope and donor
binding energy parameters, which is beyond the scope of this study.

\begin{figure}        
\includegraphics[width=\columnwidth]{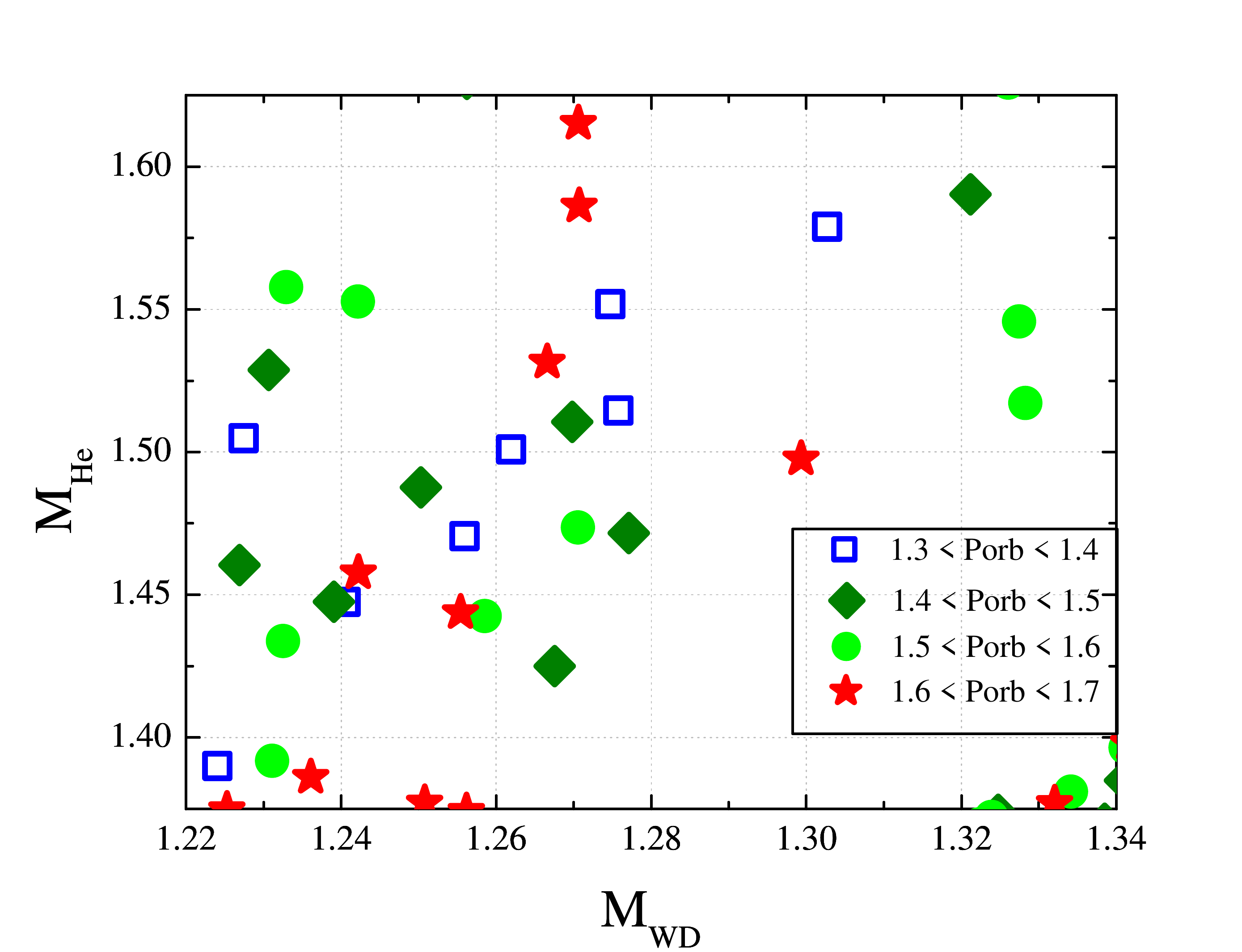}
\caption{Masses of components and periods in model systems similar to \hd. The vertical axis corresponds to subdwarf mass, and horizontal -- to the mass of  WD. Only systems with He-stars born less than in 5~Myr after WD are shown. 
The colors and symbols indicate the orbital period given in days (see the legend). Calculations are done for the common envelope parameter $\alpha_\mathrm{ce}=2$.  }
\label{fig:m1_m2_hd}
\end{figure}                  
 
In the simulation, we assumed that the primary components follow   
Salpeter IMF and we used  flat initial distributions for the mass ratios of components and for the logarithm of orbital separation. Then, if there are $10^{11}$ stars in the Galaxy
and the  binarity rate is about 50\%  \citep[see][Appendix A]{2013A&A...552A..69V}, we roughly estimate that,
currently, in the Galaxy exist $\sim 25$  \hd-like systems in which a He-star forms within 5~Myr after the
WD and have masses of components and period within $\pm0.1\,\ms$ and
$\pm0.2$\, day of the  observed values, respectively. These systems are shown in Fig.~\ref{fig:m1_m2_hd}.
A slight relaxation of the required binary parameters ($\pm0.2\,\ms$ for component masses) resulted in $\sim500$ systems in the Galaxy, so that the discovery of one of them within 650 pc is quite probable. 

\section{Discussion}
\label{sec:disc}

Our hypothesis of a young WD still in the contracting phase can solve the puzzle of the spin-up of the pulsar companion of \hd . As extensively discussed in \citet{2016MNRAS.458.3523M}, such a  spin-up is difficult to explain in a system where the compact object accretes matter  at the low rate that can be provided by the tenuous wind of the sdO mass donor. An accretion rate able to provide enough angular momentum for a WD,  would also yield a large luminosity, implying that the commonly adopted distance of \hd\  (650 pc, \citealt{1978A&A....70..653K}) has been   underestimated by a factor of ten or more, which seems very unlikely.  In the case of a NS, the luminosity would fit the observations, but an unusually low magnetic field would be required. 


On the other hand, if the spin-up is caused by the secular decrease of the moment of inertia in a young contracting WD,  we obtain the correct value of $\pdot$  for a reasonable range of masses and ages consistent, respectively,  with the measured values and with the evolution of this binary.
The model we used to calculate the WD evolutionary sequence is based on a set of robust assumptions. Although it does not include some refinements used in the most up-to-date models of WD evolution, this is not crucial for the purposes of this study. This is supported 
by recent works in which a similar technique is used to constrain the neutrino emission of hot WDs., e.g. \citep{2014A&A...562A.123M,2015ApJ...809..141H}, and references therein. 

Secular spin-up,  with $\pdot$ in the range $\sim5\times10^{-13}$~--~$10^{-10}$ s s$^{-1}$, has been detected in about ten WDs in binary systems of the intermediate polar type (see, e.g. the recent compilation in \citealt{2017MNRAS.467..428D}). These WDs have magnetic fields of about 1-20 MG and  accrete from  main sequence or evolved subgiant companions that are filling their Roche-lobe. Accretion proceeds through a disk which is truncated at an inner radius, determined by the balance between the magnetic pressure and the ram pressure of the inflowing mass. The inner disk radius is  larger by a factor from tens to hundreds than the WD radius. This is very different from the case of \hd\ binary, where the mass donor is well within its Roche-lobe\footnote{The  Roche-lobe radius is $\sim3\rsun$, while the radius of \hd\ is  1.45$\pm$0.25 $\rsun$ \citep{1978A&A....70..653K}.} and the WD is accreting from the stellar wind. The spin-up rates observed in intermediate polars are fully consistent with those expected from the transfer of angular momentum to the WD caused by the mass accretion through a disk, contrary to what occurs in our system. In fact, as shown in Section \ref{sec_pdot}, the small value of the accretion rate $\dot M$ and the short spin period  imply a maximum   spin-up rate two orders of magnitude below the observed value (even in the most favorable condition, i.e. an accretion disk truncated at the corotation radius).

In our interpretation,  accretion at the current rate does not significantly influence the spin period evolution.  Accurate measurements of the period behavior and luminosity can help to test our hypothesis. In fact, we predict that small luminosity variations should not be accompanied by changes in $\pdot$. Only in the case of a major luminosity increase, which is unlikely to occur given the properties of the stellar wind of \hd , we would expect a noticeable effect on the spin period derivative.

As in our model the spin-up does not depend on the magnetic field of the WD, we cannot use the $\dot P$ value to estimate it. The only limitation comes from the evidence of stable accretion, which requires the Alfven radius, $R_\mathrm{A}=(\mu^2/(\dot M\sqrt{2G M}))^{2/7}$, to be smaller than the corotation radius.
Here $\mu=BR^3$ is the magnetic moment of the WD ($B$ is the field at the equator).  
Thus, assuming $R_\mathrm{A}= R_\mathrm{co}$ the estimate of $\mu$ is:
\begin{equation}
\mu=2^{1/4}(GM)^{5/6}\dot M^{1/2}\omega^{-7/6}.
\end{equation}
With $\dot M=3\times 10^{14}$~g~s$^{-1}$  we obtain $\mu \sim 10^{29.5}$~G~cm$^3$  and $B\sim 10^4$~G.  As for accretion it is necessary to have $R_\mathrm{A}< R_\mathrm{co}$, we obtained a rough upper limit for $B$, so the field is about few kG.
Note that $R_\mathrm{A}$, derived here under the assumption of a dipolar field, is only a factor of a few larger than the WD radius, meaning that the magnetosphere is squeezed close to the star. More realistically the magnetic field might have a complex geometry, with multipolar components able to channel the accretion flow in a hot spot much smaller than the WD radius, as required to explain the large pulsed fraction  ($\sim$65\%) and  emitting radius of the thermal X-ray emission \citep{2016MNRAS.458.3523M}.
 
We followed the evolution of WDs up to ages comparable to the Galactic lifetime. After $\sim 10^8$~yrs $\dot P$ starts to decrease faster (approximately as $t^{-1.5}$), and reaches values $\lesssim 10^{-19}$ at the age $\sim 5$~Gyrs. So, the WD cannot spin-up significantly in its future due to contraction. Thus, at some point the spin evolution of the \hd\ companion will be driven by the angular momentum transfered through accretion.   

We can also estimate the initial period of the WD taking its present day value 
of 13.2 s 
and the age of 2 Myrs. Assuming that the spin period did not change much during the relatively short common envelope stage, the initial value is $P_0=P (I_0/I)\sim 4$~min, where $P$ and $I$ are the current values. 
So, the compact object has been significantly spun-up during its lifetime
due to contraction.



The predicted value of $\ddot P$ due to the WD contraction is very small. It is comparable to the smallest values of the second period derivatives in radio pulsars. Thus, it would be very difficult to measure it with X-ray observations. 



\begin{figure}
\centering
\includegraphics[width=\linewidth]{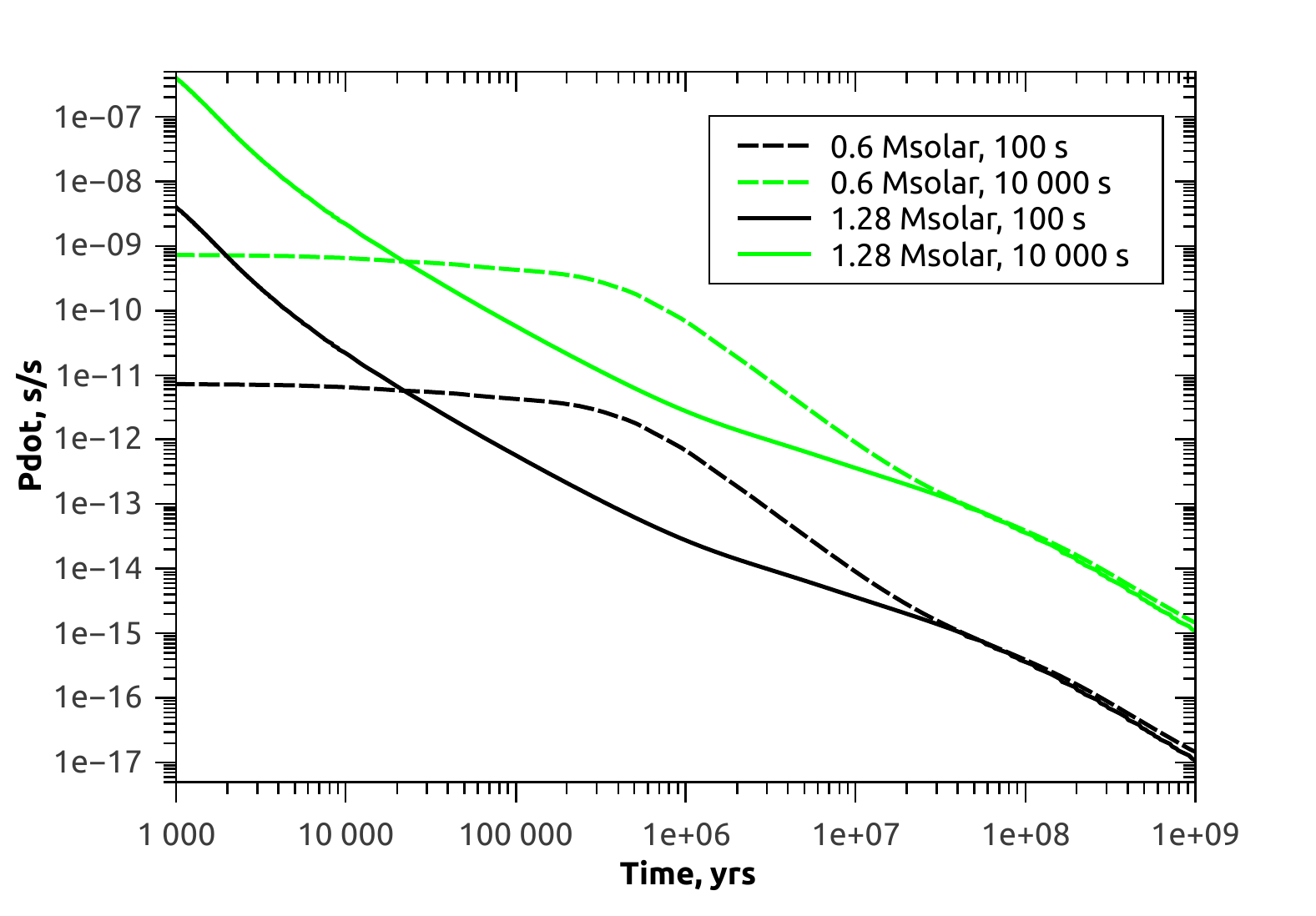}
\caption{Evolution of $\dot P$. Solid lines correspond to $M=1.28\, M_\odot$. Dashed lines to $0.6\, M_\odot$. Two upper (lighter and thicker, green in electronic version) lines are plotted for spin period $10^4$~s, and two lower for $P=100$~s. }
\label{pdot2}
\end{figure}

As $\dot P\propto P$ (see eq. 3) and rotation does not influence the internal structure significantly (i.e.,  $I$ does not depend on $P$) we expect that in similar sources with young accreting WDs with more typical spin periods, the  values of $\dot  P$ can be larger. 
As an example, we show the evolution of $\dot P$ vs. age for two values of $P$ and two WD masses in Fig.~\ref{pdot2}. It is seen that WDs with a typical mass $0.6\, M_\odot$ at ages $\sim$~few hundred thousand years can have very large period derivatives, especially for large spin periods. It is possible that the process described here is at work also in some of the cataclysmic variable which show a secular spin-up, provided the WDs are sufficiently young. Unfortunately, this is difficult to demonstrate due to the presence of significant  accretion torques, which by themselves are already able to account for the observed spin-up rates. It would be important to look for other low-luminosity X-ray pulsars with WDs with ages $\lesssim 10^8$ yrs, similar to \hr . Note that the low  accretion rate   in this system is due to the particular nature of the mass donor, i.e. a hot subdwarf fitting inside the Roche-lobe, but endowed with a weak stellar wind. This yields a small luminosity and a very soft X-ray spectrum, properties which unfortunately hamper the detection of similar systems. Future X-ray facilities, especially all-sky surveys such as the one planned with eRosita, will hopefully provide more candidates to test our proposed scenario and further investigate the early stages of WD evolution.

\section{Summary}\label{sec:summary}

In this paper we have proposed a novel interpretation which, contrary to explanations related to mass accretion,  can naturally explain the  spin-up  of the compact object in the peculiar X-ray binary \hr . 
We showed that the contraction of a WD with mass of $1.28\, M_\odot$  and  age of about 2 Myrs can produce the observed spin-up rate of  $\dot P = 2.15\times 10^{-15}$ s s$^{-1}$.

If our hypothesis is correct, it could be the first direct observational evidence of a young contracting WD and gives us
the unique opportunity to probe early stages of WD evolution. 

\section*{Acknowledgements}
 S.B.P. and A.G.K. acknowledge support from RSF grant No. 14-12-00146.
 S.I.B. is grateful to W.~Hillebrandt and H.-T.~Janka for support of the 
visit to MPA where part of this work was done.
LRY was supported by Presidium of RAS Basic Research Program P-41.
We also would like to thank the organizers of the conference ``Physics of neutron stars -- 2017'' in St.Petersburg, 
where we started to discuss this problem.
This research has made use of NASA's Astrophysics Data System.




\bibliographystyle{mnras}
\bibliography{wd}

\appendix
\section{Code validation}
\label{sec:code}

To calculate the WD cooling we used the code described by  \cite{1993ARep...37..187B, 1994MNRAS.266..289B}.
In this Appendix we compare the results of our code with those obtained with a couple of more recent simulations of WD cooling.

\subsection{Comparison with Bergeron's code}

First, we use models from P.~Bergeron's
website\footnote{\url{http://www.astro.umontreal.ca/~bergeron/CoolingModels/}}.
Some of those models are described by \cite{2001PASP..113..409F},
see new details in \cite{2011ApJ...737...28B}.

We have used the evolutionary sequence {\tt CO\_1200204},
which corresponds to a mass $M=1.2 M_\odot$ and initially ``thick'' H and He layers with $q_{\rm H}=10^{-4}$ and $q_{\rm He}=10^{-2}$,
where $q$ is a fraction of the total mass.
It has a mixed C/O core composition (50/50 by mass fraction mixed uniformly).

For our code we used a very similar model with $M=1.2 M_\odot$, with outer layers $M_{\rm H}=1.4 \times 10^{-4} M_\odot$
and  $M_{\rm He}=2.6 \times 10^{-2} M_\odot$ (the same H and He envelopes we used for our models of the WD companion of \hd ) .


\begin{figure}
\centering
\includegraphics[width=\linewidth]{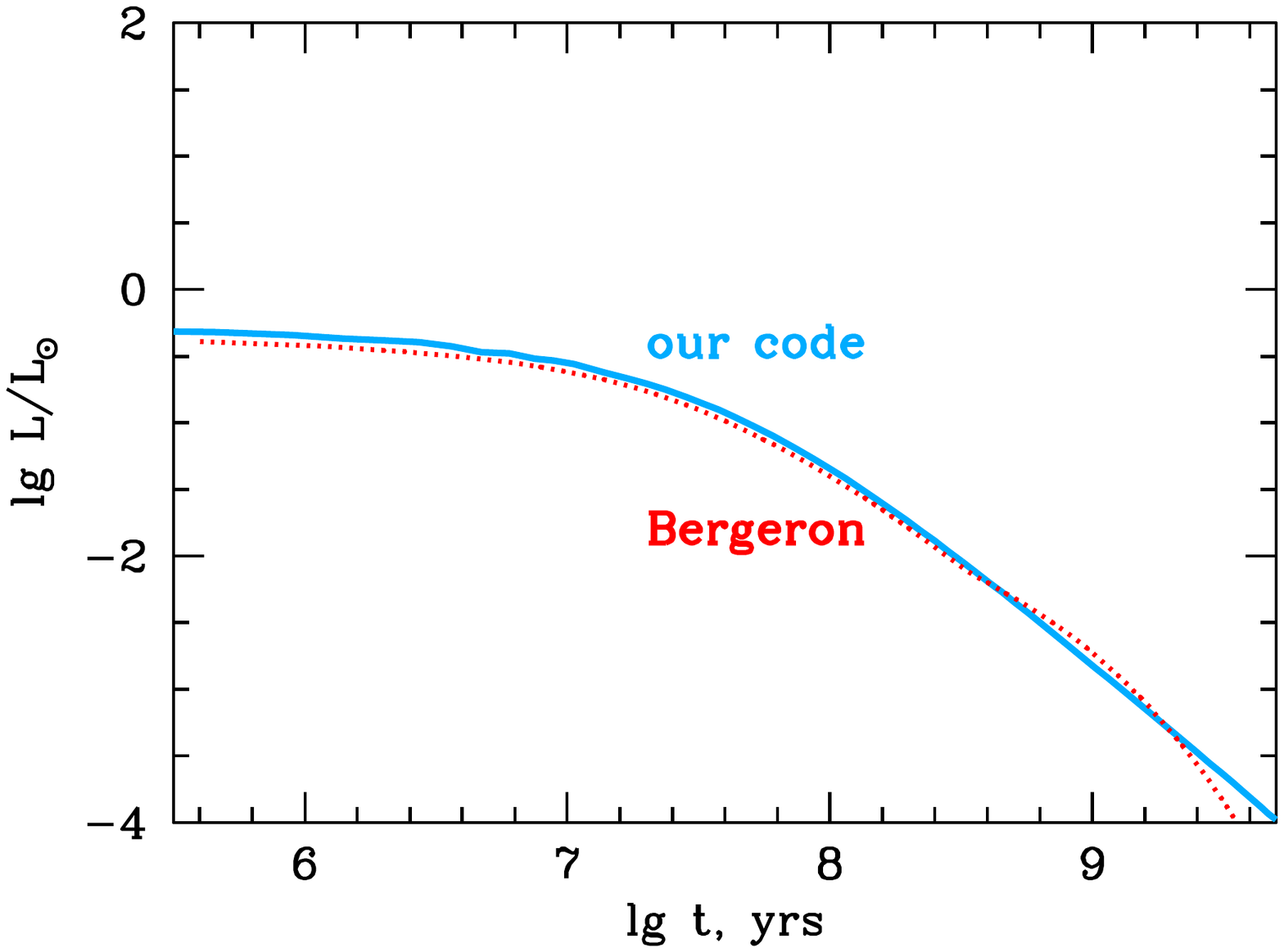}
\caption{Evolution of luminosity $L$ of a $1.2 M_\odot$ WD according to our code (thick blue solid line) and that of Bergeron et al. (thin red dotted line). }
\label{fig_bergL}
\end{figure}

Since zero epochs in both simulations are arbitrary, we plot here our results shifting them slightly along the time axis, and starting
at the moment closest to the first Bergeron's output, when his $T_{\rm eff}=59280$ K.

\begin{figure}
\centering
\includegraphics[width=\linewidth]{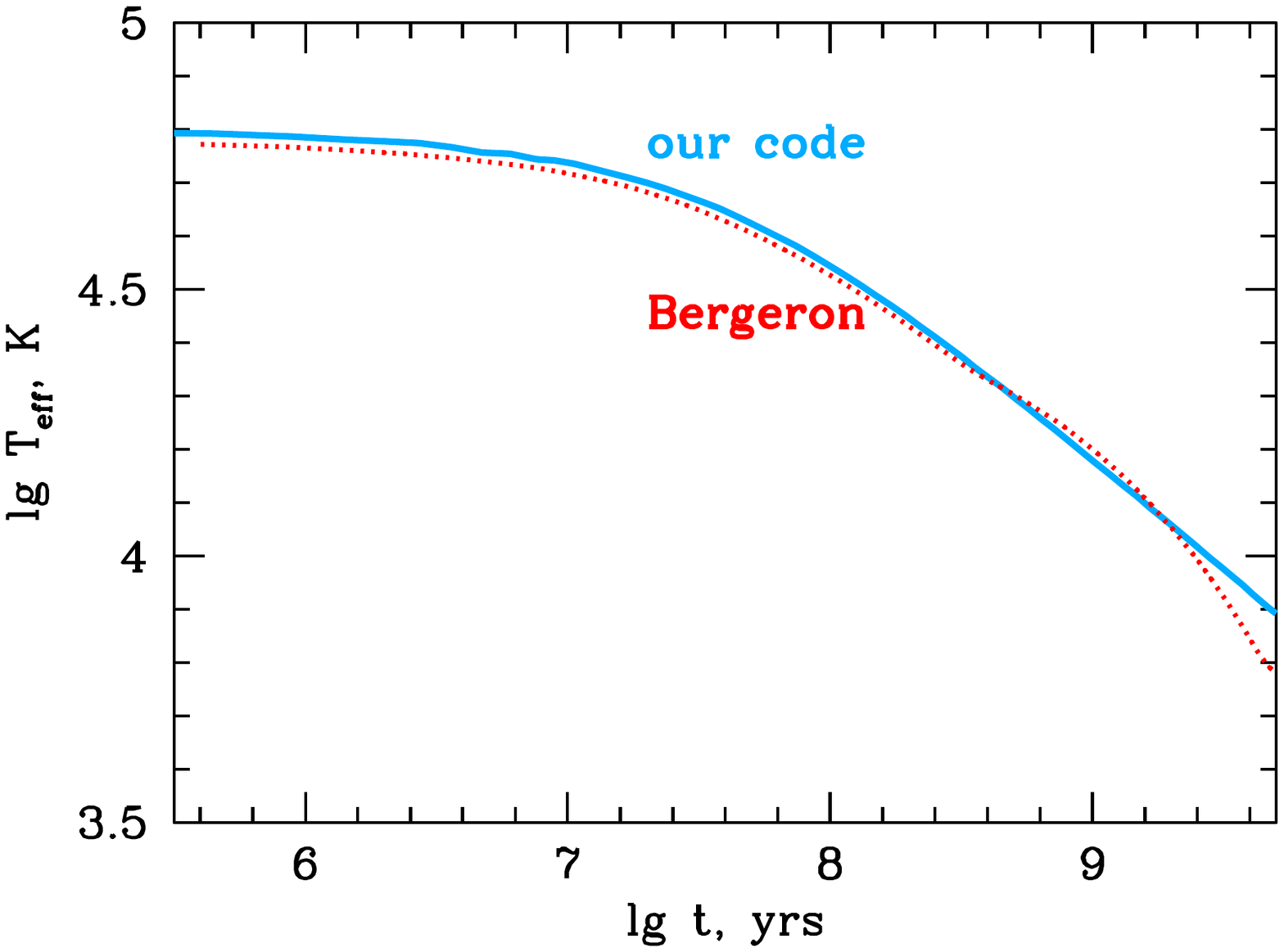}
\caption{Evolution of effective temperature $T_{\rm eff}$ of a $1.2 M_\odot$ WD according to our code (thick blue solid line) and that of Bergeron et al. (thin red dotted line). }
\label{fig_bergT}
\end{figure}

Figure~\ref{fig_bergL} demonstrates the agreement of luminosity, and Fig.~\ref{fig_bergT} that of effective temperature between the two codes.
As described in \cite{1994MNRAS.266..289B}, our code has relevant physics when a WD is hot enough,
$T_{\rm eff}>1.2 \times 10^4$.  Nevertheless our curves are in reasonable agreement with Bergeron's code even at late epochs.

\begin{figure}
\centering
\includegraphics[width=\linewidth]{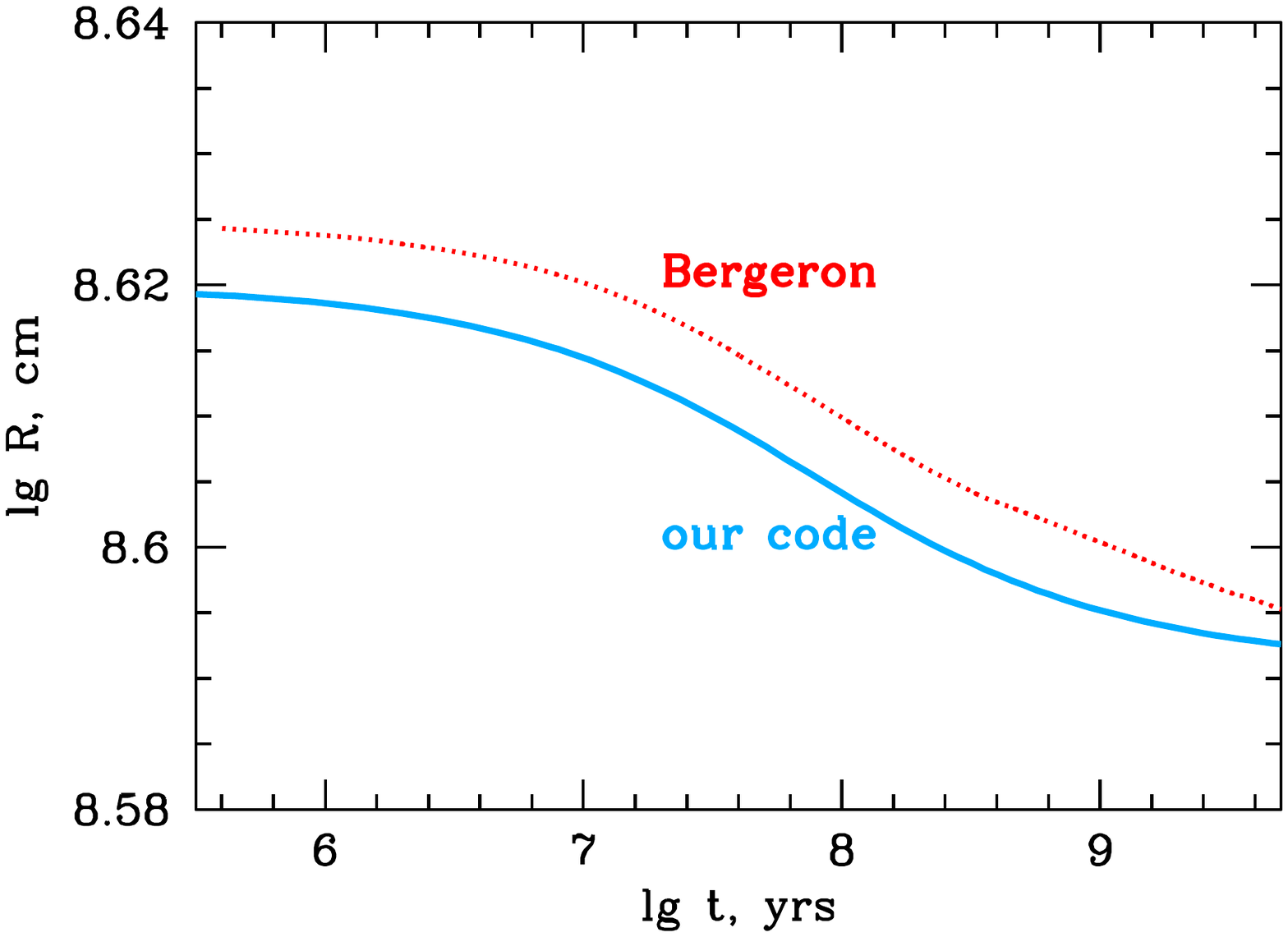}
\caption{Evolution of radius $R$ of a $1.2 M_\odot$ WD according to our code (thick blue solid line) and that of Bergeron et al. (thin red dotted line). }
\label{fig_bergR}
\end{figure}

There is a tiny difference in radii in Fig.~\ref{fig_bergR} in comparison with our results. This small discrepancy,   at the level of 1\%,  is probably caused by different masses of H and He envelopes, chemical composition, equations of state, etc., but the behavior of $R(t)$ is the same in both cases. Thus, this difference does not significantly influence our conclusions.

\subsection{Comparison with BaSTI code}

The same $1.2 M_\odot$ model computed with our code has been compared with another independent simulation done by \cite{2013A&A...555A..96S}, who used the BaSTI code\footnote{A Bag of Stellar Tracks and Isochrones, see  \url{http://basti.oa-teramo.inaf.it/index.html}}.

We have used tables at the BaSTI website  which contain data on the $L$ and $T_{\rm eff}$ evolution for a $1.2 M_\odot$ WD.
The specific model is {\tt COOL120BaSTIfinaleDAnosep}, which refers to a  $M=1.2 M_\odot$ DA white dwarf without separation of phases at ion crystallization stages.

\begin{figure}
\centering
\includegraphics[width=\linewidth]{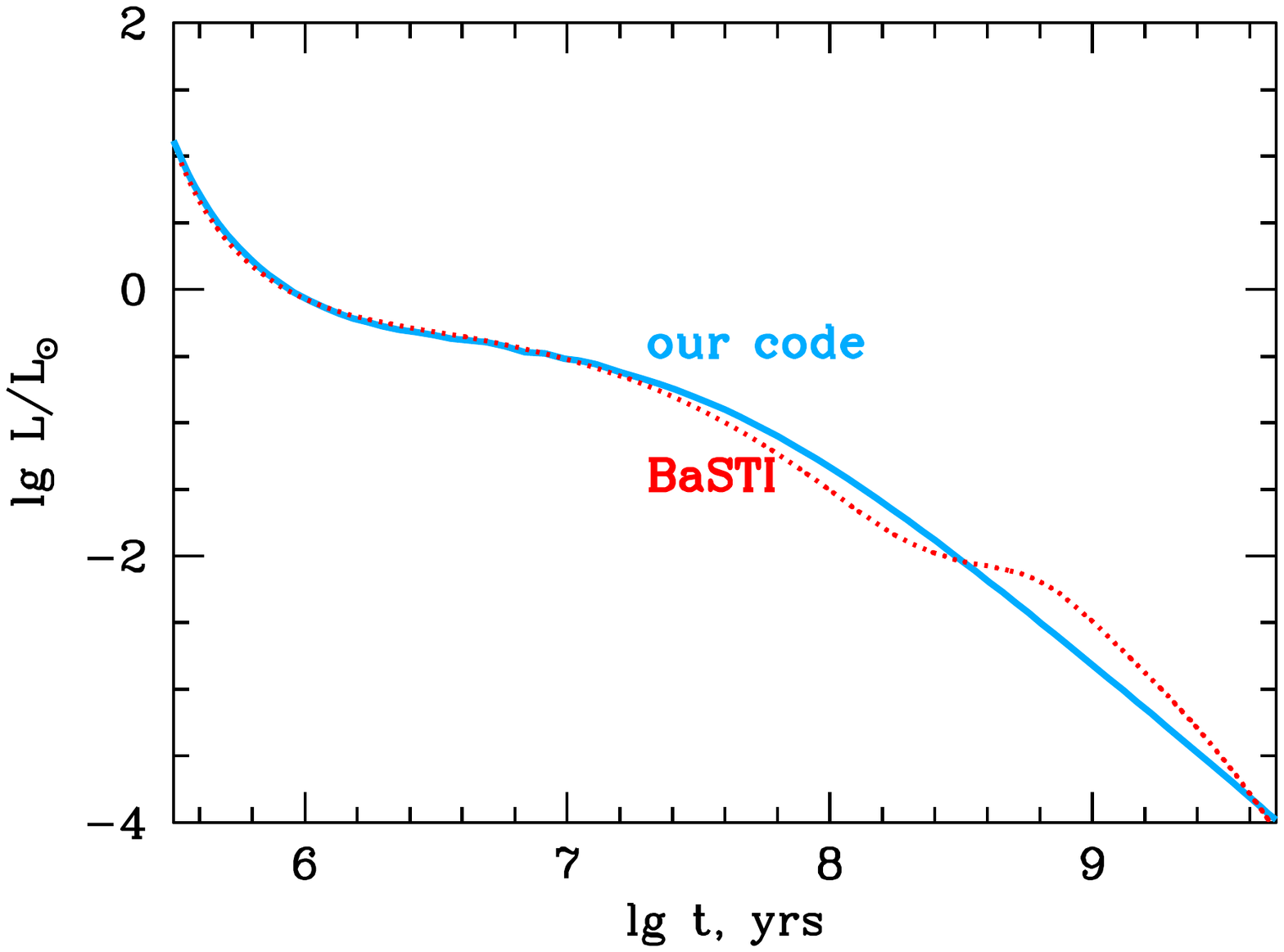}
\caption{Long time scale evolution of luminosity $L$ of a $1.2 M_\odot$ WD according to our code (thick blue solid line) and that of BaSTI (thin red dotted line). }
\label{fig_bastiLl}
\end{figure}

Since the data on the BaSTI website contain earlier epochs we have shifted our output to higher temperatures, accordingly.

\begin{figure}
\centering
\includegraphics[width=\linewidth]{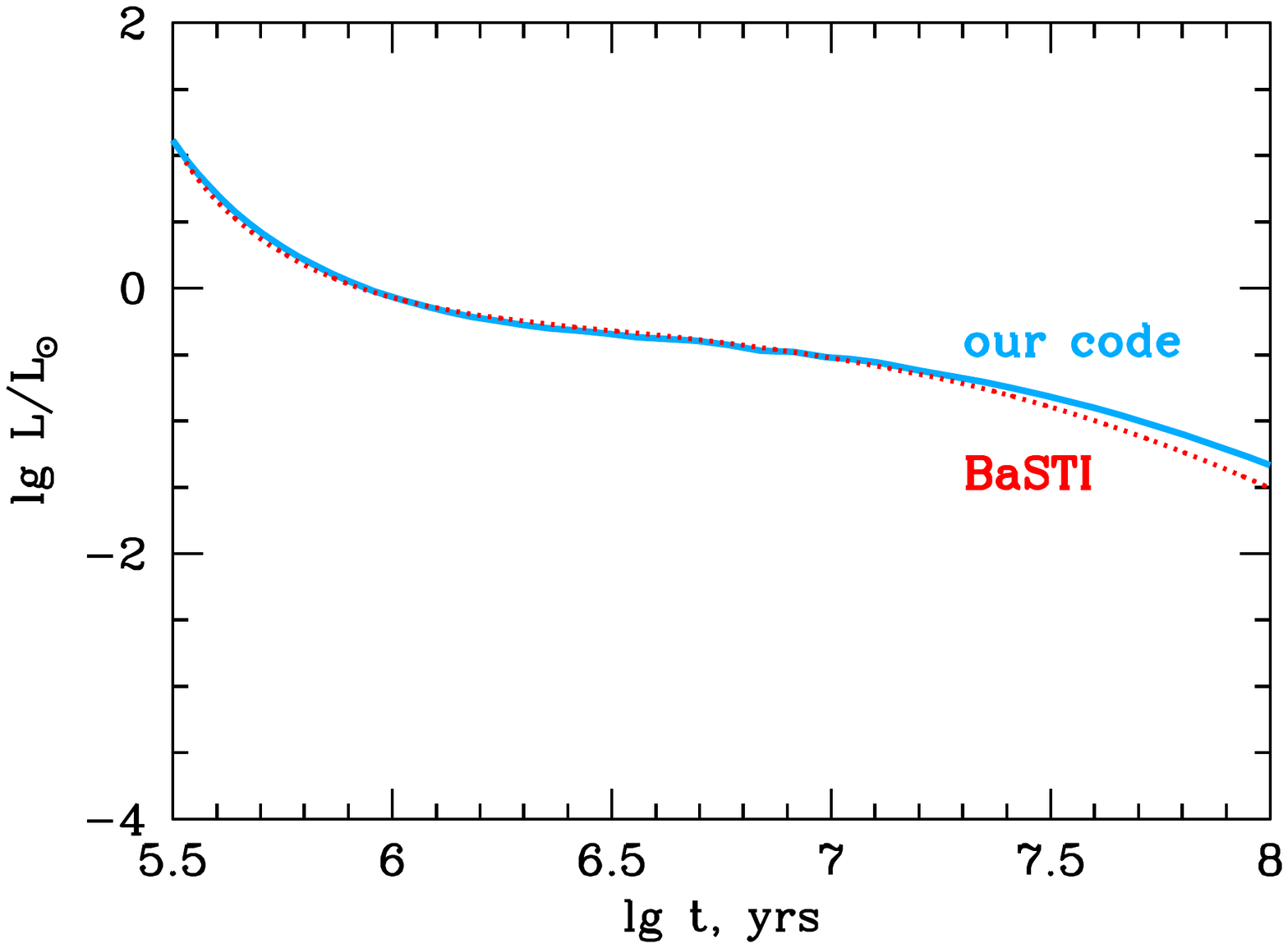}
\caption{Evolution of luminosity $L$ of a $1.2 M_\odot$ WD according to our code (thick blue solid  line) and that of BaSTI (thin red dotted line) on  short time scale. }
\label{fig_bastiLs}
\end{figure}

\begin{figure}
\centering
\includegraphics[width=\linewidth]{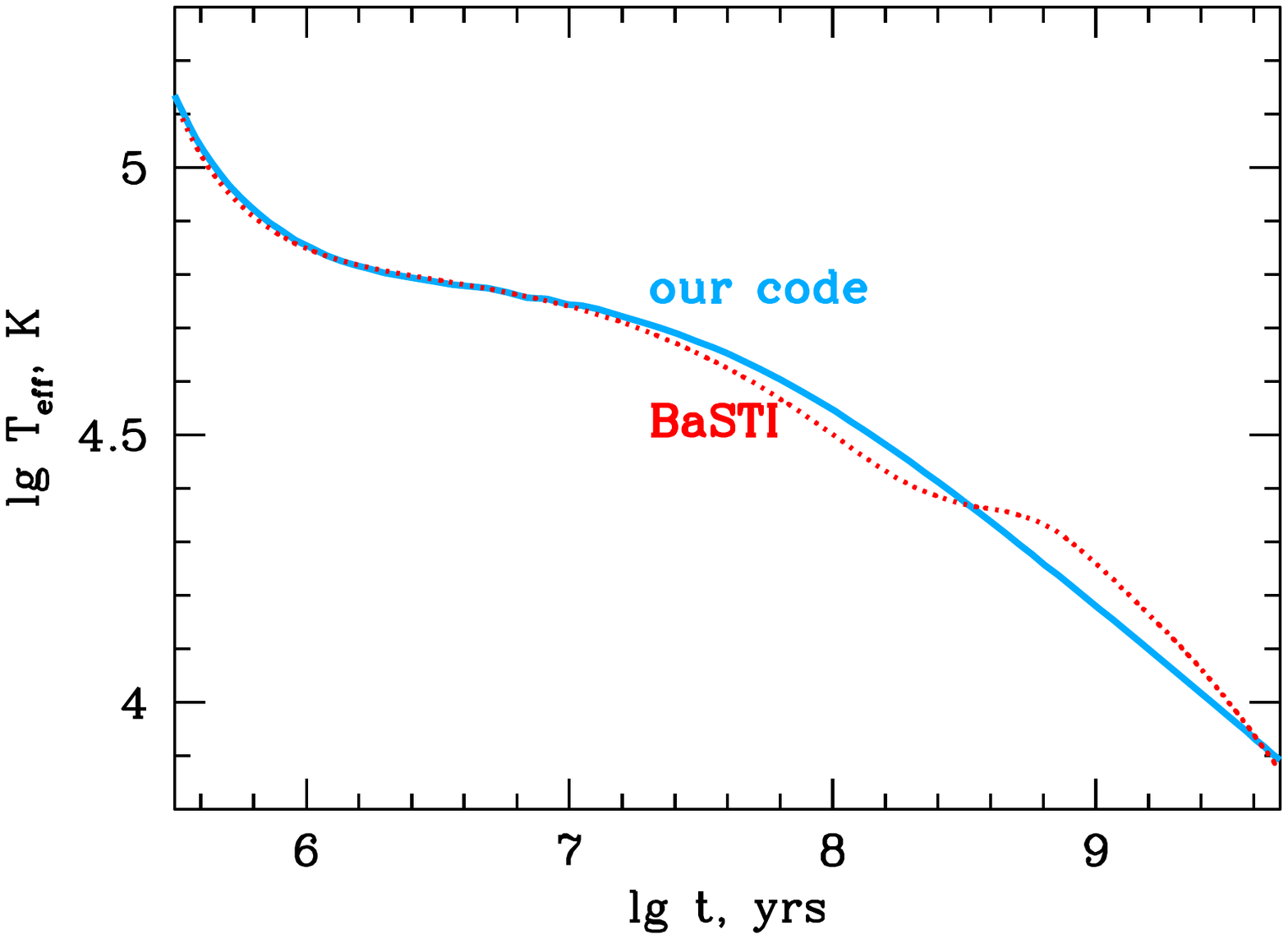}
\caption{Long time scale evolution of effective temperature $T_{\rm eff}$ of a $1.2 M_\odot$ WD according to our code (thick blue solid line) and that of BaSTI (thin red dotted line). }
\label{fig_bastiTl}
\end{figure}

\begin{figure}
\centering
\includegraphics[width=\linewidth]{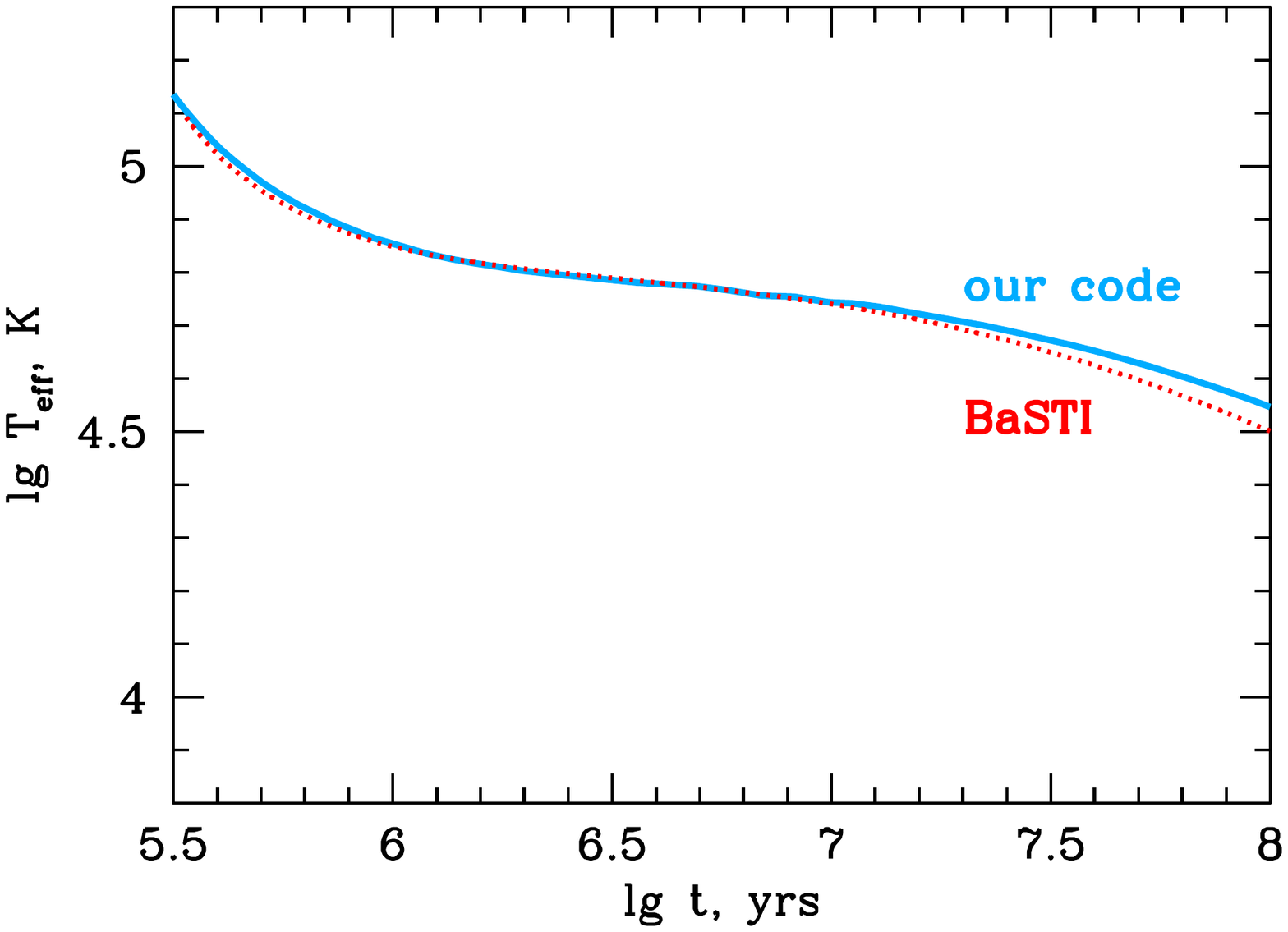}
\caption{Evolution of effective temperature $T_{\rm eff}$ of a $1.2 M_\odot$ WD according to our code (thick blue dashed line) and that of BaSTI (thin red dotted line) on the short age scale. }
\label{fig_bastiTs}
\end{figure}

On late stages (Figs.~\ref{fig_bastiLl}, \ref{fig_bastiTl}) the discrepancy is rather large (since our code is developed only for hot WDs), but for epochs which are
interesting for us (Figs.~\ref{fig_bastiLs},  \ref{fig_bastiTs}) in this paper the agreement of our  results with this modern code is just perfect.

We conclude that there are no doubts on the validity of the approximations used in our approach to model the WD in the binary system \hr.


\end{document}